\documentclass{aastex}
\usepackage{emulateapj5}

\newcommand{\CXO}{{\it CXO}\ }
\newcommand{\ACIS}{{\it ACIS}\ }
\newcommand{\ROSAT}{{\it ROSAT}\ }
\newcommand{\HRI}{{\it HRI}\ }
\newcommand{\HST}{{\it HST}\ }

\newcommand{\psr}{\ensuremath{\rm PSR\:J0538+2817}\ }

\newcommand{\et}{{\it et~al.}}

\newenvironment{inlinefigure}{
\smallskip
\def\@captype{figure}
\noindent\begin{minipage}{0.999\linewidth}\begin{center}}
{\end{center}\end{minipage}\smallskip}

\slugcomment{\apj \, submitted}
\shorttitle{PWN torus of \psr}
\shortauthors{Romani \& Ng}

\begin{document}

\title{The PWN torus of \psr and the Origin of Pulsar Velocities }

\author{Roger W. Romani \& C.-Y. Ng}
\affil{Department of Physics, Stanford University, Stanford, CA 94305}
\email{rwr@astro.stanford.edu, ncy@astro.stanford.edu}

\begin{abstract}

	We find evidence for a faint wind nebula surrounding \psr
in \CXO-\ACIS imaging. This object is particularly interesting, as
the pulsar spindown age is largest for any such X-ray PWN.
If interpreted as an equatorial torus, the PWN supports the claimed association
with the S147 supernova remnant and implies good alignment between the pulsar spin
and space velocity. Comparison of the SNR age, X-ray cooling age
and characteristic age suggests a birth spin period of $\ga 130$ms.
In turn, if we accept as causal the alignment of the linear and angular
momenta, this places strong constraints on the origin of the
`kick' at the neutron star birth.
\end{abstract}

\keywords{stars: neutron --  pulsars -- stars: rotation}

\section{Introduction}

	{\it Hubble Space Telescope} and, especially, {\it Chandra X-ray
Observatory} (\CXO) imaging have discovered toroidal
structure in the synchrotron pulsar wind nebulae (PWN) of relativistic
particles and magnetic fields around
young neutron stars (e.g. Weisskopf \et\, 2000, Pavlov \et\, 2001, Gotthelf 2001).
Of these, the Crab and Vela pulsar tori are the most striking, 
with clear symmetry (pulsar spin) axes. 
Moreover, for these two objects \HST proper motion
studies show alignment between the position angles of the
projected ${\vec \Omega}$ and ${\vec v}$. The chance probability of two such
alignments is modest, but if taken as causal this gives strong
constraints on nature of the `kick' at pulsar birth which typically produces
a large space velocity (Spruit \& Phinney 1998,
Lai \et\, 2001). Even a few further examples could improve the statistical
significance of the  ${\vec \Omega}$-${\vec v}$ correlation and further
constrain the kick physics (\S 4). The challenge is to identify sources for
which both PWN tori and 2-D proper motions are measurable; after the Crab
and Vela, many bright young PSR/PWN sources are quite distant and 
direct proper motion measurement is challenging. We 
report here on \CXO imaging of one particularly interesting example.

	\psr, discovered by \citet{a96} is a 143ms pulsar with
characteristic age $\tau_c = 6 \times 10^5$y. Located within the
old $\sim 10^5$y SNR S147 (G180.0$-$1.7)  with a DM-estimated
distance of 1.2kpc \citep{cl02} comparable to that of the SNR, 
it is believed to be a likely association despite the age discrepancy.
Previously reported X-ray observations are limited to \ROSAT all sky survey 
(Sun et al 1996) and \HRI\, imaging.  Inspection of the
archival {\it HRI} data showed a source slightly extended beyond the 
nominal PSF, although aspect errors may have contributed. We have obtained
\ACIS-S imaging to resolve this possible PWN.

\section{\CXO Observations}

	We observed \psr with the \CXO {\it ACIS} on February 7-8, 2002,
with the pulsar near the S3 chip aim-point, obtaining 19.5ks of TE
exposure.  All 6 \ACIS chips were active and several field
sources were detected off axis.  None showed markedly unusual behavior;
we therefore concentrate on the S3 pulsar data.  Calibration, processing 
and spectral fitting were performed with CIAO 2.3, including the latest
calibrations and CTE corrections.

\subsection{Spatial Decomposition}

	The pulsar core produces 0.13 cnts/s and, with the standard 3.2s ACIS
frame, suffers 20\% pileup.  In Figure 1 we compare the 
0.5-5\,keV source radial profile with that of the HRMA+ACIS model PSF (generated 
for the observed source count spectrum and corrected for pileup).
The point source normalization has been obtained from a PSF fit to the central 
3$^{\prime\prime}$. The core is resolved, with $\sim 10$\% of the
counts in extended emission. If the PSF is convolved with a 
$\sigma=1^{\prime\prime}$ Gaussian, the counts are well matched out to
5$^{\prime\prime}$, but a $5\sigma$ excess remains above background from
$5-8^{\prime\prime}$. We thus believe that the extended emission cannot be 
attributed to aspect errors or poor PSF modeling.


\begin{inlinefigure}
\scalebox{.8}{\rotatebox{0}{\plotone{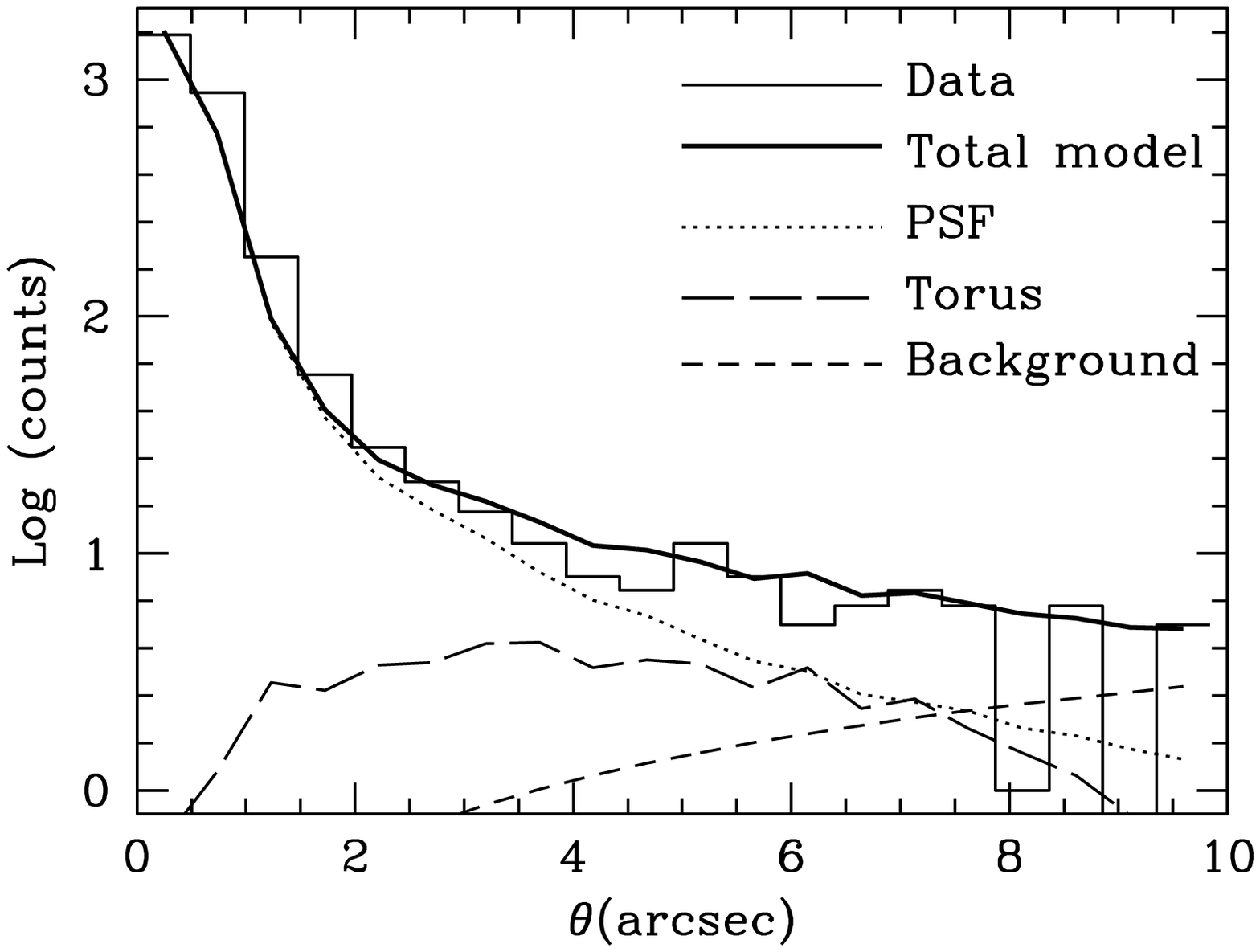}}}
\figcaption{Source and model radial count distributions.
\label{radprof}}
\end{inlinefigure}


	This emission does not follow the symmetric PSF wings (Figure 2). 
We start by fitting elliptical isophotes, with 
semi-major axes from 1-10$^{\prime\prime}$,
using the STSDAS routine {\it ellipse}. For small $1-2.5^{\prime\prime}$
semi-major axes, we find an ellipticity of $\sim 0.1-0.2$ at position angle
$\sim 130^\circ$ (measured N-E). From $4-9^{\prime\prime}$ the PA changes to a
weighted average value of 61.0$\pm 5.5 ^\circ$ with an ellipticity of 0.25-0.35.
In Figure 2, the core has a tail to the SE defining the inner ellipse, while
the scattered photons along the NE-SW axis represent the larger ellipse. 
Some trailing along the read-out direction 
is visible in the detector coordinate image (at PA= 91$^\circ$), but cannot
be seen in Figure 2. The measured trail flux contributes less than 1 count 
in the $10^{\prime\prime}$ region of our extended emission, and does not
lie along the axis of extension.
 
	In analogy to other pulsar/PWN systems we can expect a toroidal wind
shock surrounding the pulsar, possibly with a polar jet. 
The $4-9^{\prime\prime}$ ellipse has a scale comparable to the expected
wind shock radius for a confinement pressure appropriate to
the interior of S147 (see \S 3). We therefore adopt this physical model,
computing the expected photons from the point source PSF wings and
fitting the excess photons outside of $3^{\prime\prime}$ to an
inclined torus.  A uniform background for the residual
counts is also allowed.  The torus model is characterized by a 
radius $r_T$, a Gaussian blur torus thickness $\delta$, the position angle $\Psi$ 
of the torus axis and its inclination $\zeta$ from the observer line of sight.
We assume that the post-shock flow remains relativistic, so 
that the emission is Doppler boosted in the forward direction. For a rest
power law emission spectrum 
$dN^\prime_\gamma/dE_\gamma \propto E_\gamma^{-\alpha}$,
we expect an observed emissivity
$$
dN_\gamma/dE_\gamma \propto (1- {\bf n} \cdot \beta)^{-(1+\alpha)}
 E_\gamma^{-\alpha}
$$
where $\alpha$ is the photon spectral index, $\beta = {\bf v}/c$ is the 
bulk velocity of the post shock flow and ${\bf n}$ is the unit normal toward
the observer (e.g. Pelling \et~ 1987). For typical PWNe,
$\alpha \approx 1.5$ and the emission brightens on the side
facing the observer. 

	Of the 2875 counts in the central 32$^{\prime\prime} \times
32^{\prime\prime}$ region, the best-fit point source provides 2430, 
the uniform background contributes 80 counts and the `torus' provides
50 counts.  The remaining counts represent
unresolved fine structure in the nebula core. The best-fit torus radius
is $r_T = 7.2\pm 0.7^{\prime\prime}$, the `blur' scale is 
$\delta = 1.9\pm 0.5^{\prime\prime}$. The torus $\beta$ is
$0.53\pm 0.04$, similar to values fit for other PWN.
The torus symmetry axis is located at position angle $\Psi = 154.0\pm 5.5^\circ$
(measured N through E) and $\zeta =100\pm 6^\circ$ (into the plane of the sky),
where the confidence regions have been scaled to that of $\Psi$.


\begin{inlinefigure}
\scalebox{1.}{\rotatebox{0}{\plottwo{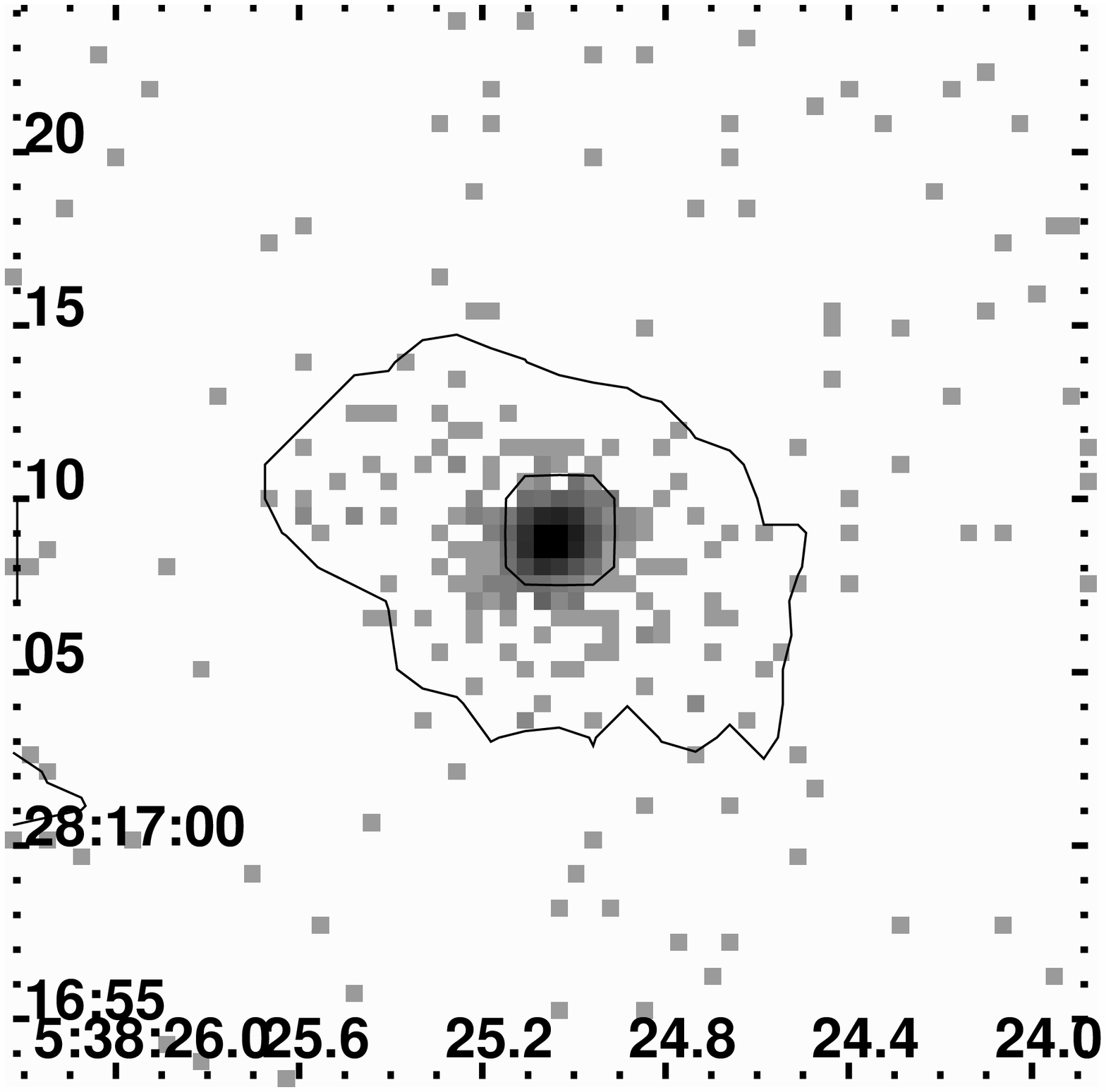}{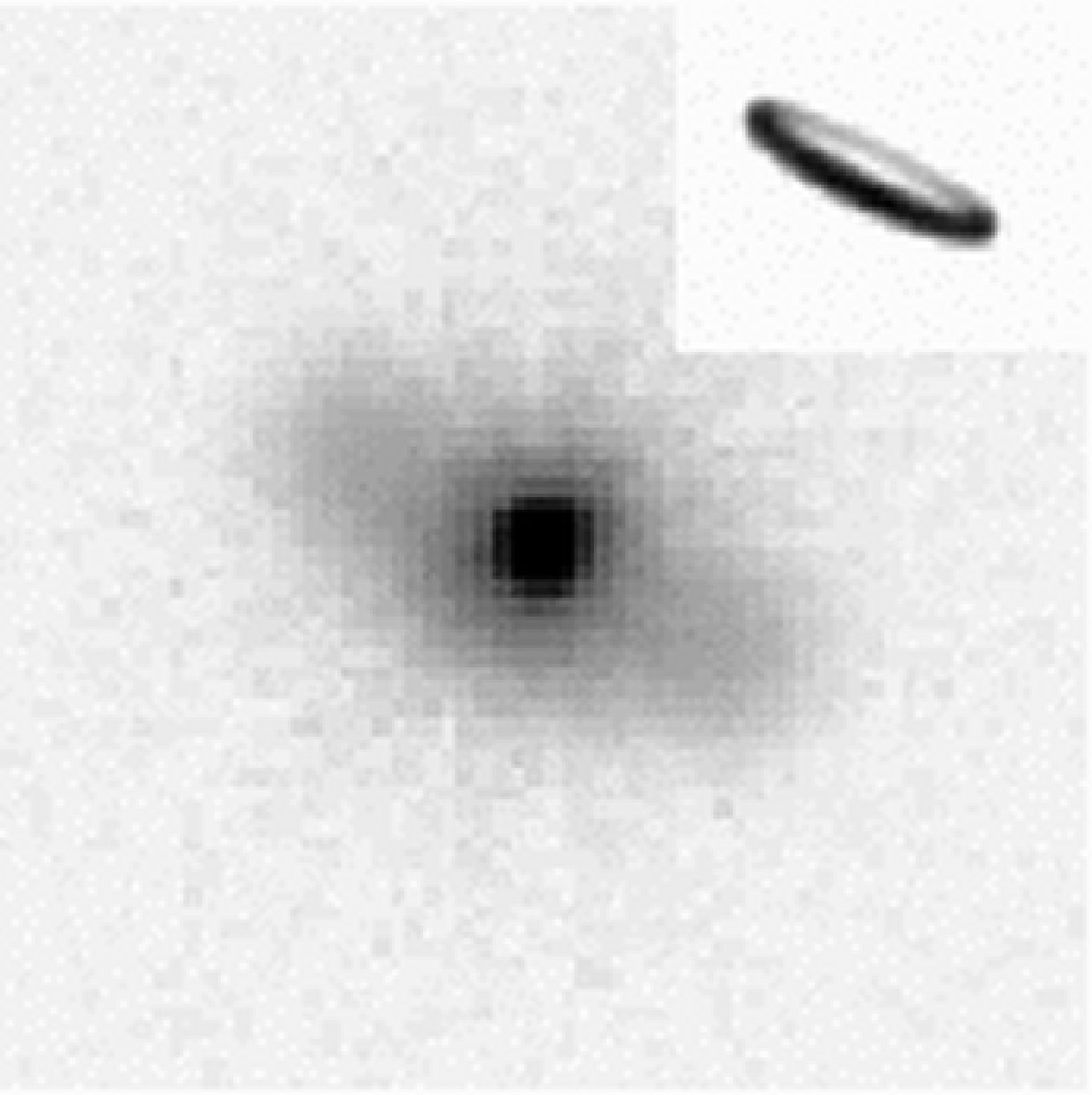}}}
\figcaption{\emph{Left:} ACIS 0.5-5\,keV image with smoothed low level contours.
\emph{Right:} Best-fit Point Source + torus model.
\emph{Inset:} Best-fit torus model at half scale, with $\delta$ reduced to 
0.5$^{\prime\prime}$ to show geometry.
\label{PWNim}}
\end{inlinefigure}


In Figure 2 we show the photon data and the predicted image for the best-fit 
point source + torus model. The position angle $\Psi$ is robust to cuts
in the fit data set and forced variations in $r_T$ and $\delta$ and agrees
well with the value determined from ellipse isophote fits; we adopt
this value as determining the symmetry axis. $\zeta$, primarily set by the
ellipse axis ratio, is also surprisingly well determined.
The brightening of the diffuse emission SE of the core indicates
the near side of the torus, enhanced by Doppler beaming
(inset). However, the core asymmetry also suggests 
a `polar jet'.  Certainly deeper exposures would be quite 
valuable in confirming the reality of these faint structures and measuring 
their precise parameters.

\subsection{Spectral Analysis}

We extract the point source from a $2^{\prime\prime}$ aperture, correcting for
pileup and measuring a nearby
background from the S3 chip.  For the torus we take a 
$\sim 10^{\prime\prime}\times 18^{\prime\prime}$ ellipse with a 
$4^{\prime\prime}$ radius exclusion core. The PSF wings of the central source
contribute significant counts in this aperture, so we include a scaled
version of the best-fit point source spectrum as a fixed background
and fit the excess to characterize the torus emission.

\begin{inlinefigure}
\scalebox{1.}{\rotatebox{0}{\plotone{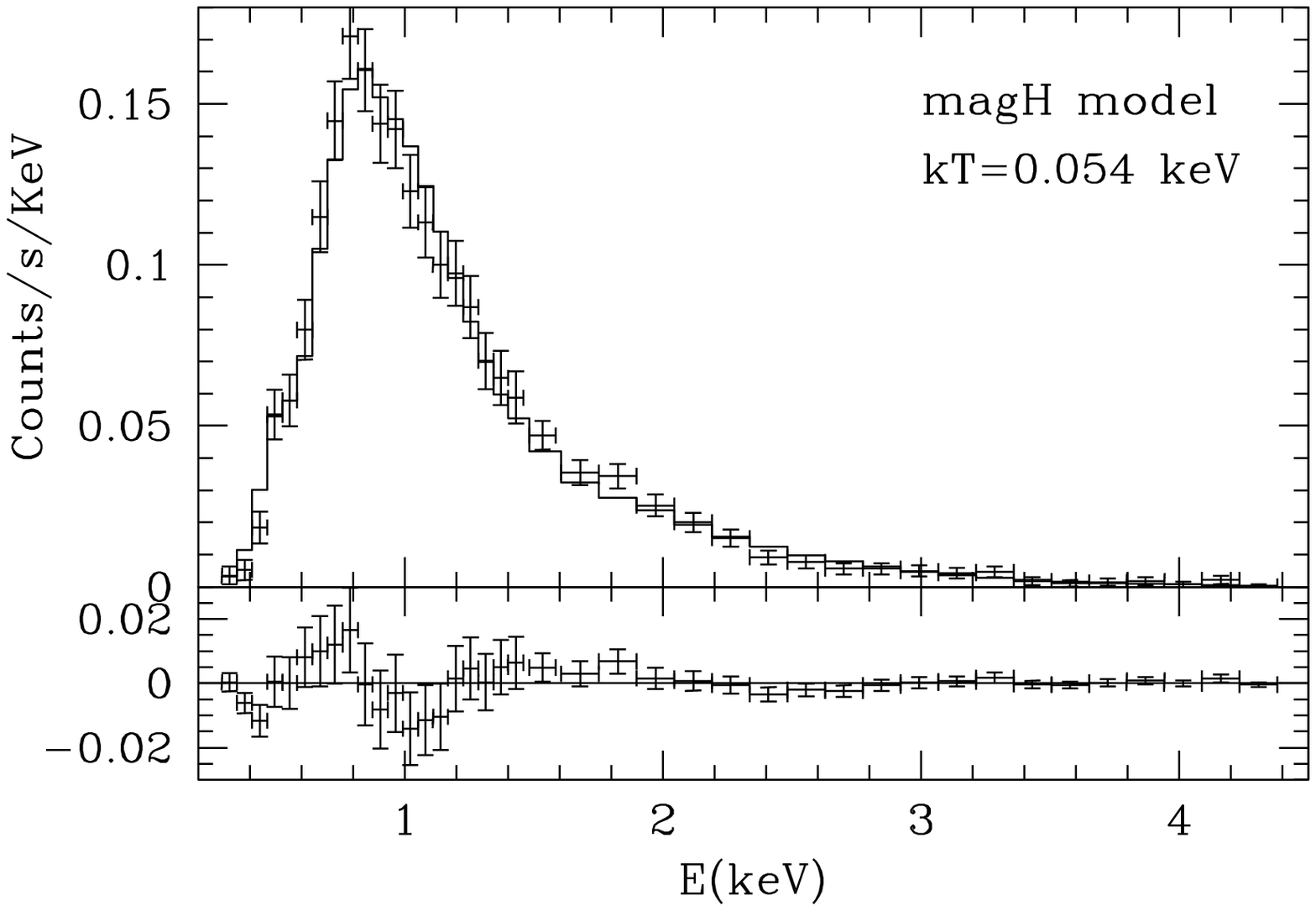}}}
\figcaption{Point source spectrum with a pileup-corrected magnetic H model 
atmosphere spectrum and residuals.
\label{spec}}
\end{inlinefigure}

The point source spectrum is adequately modeled by an absorbed blackbody,
although systematic residuals persist at low energies. Table 1 gives
the fit parameters with 95\% errors, the unabsorbed bolometric flux 
and the effective
radius of a spherical emittor at $d=1.2$kpc. The second fit is from
the magnetic H atmosphere models of \citet{pav95}, which produce
slightly better $\chi^2$ and a larger effective radius.
We note that all of these fits are somewhat suspect, since the spatial
decomposition suggests $\sim 10$\% diffuse emission even for the
core. Composite (BB+BB or BB+PL) models of course improve
the fit statistics, but are not demanded by the present data. These fits
give $N_H$ somewhat larger than the $\sim 1.2 \times 10^{21}{\rm cm^{-2}}$ 
inferred from the pulsar DM with a canonical $n_e/n_H \approx 10$.

	A comparison of the energies of the 51 torus photons outside of 
$4^{\prime\prime}$ with the 240 photons from $1-2^{\prime\prime}$ 
(which avoid pile-up in the point source core) gives a KS statistic 
probability of 0.06 that they are drawn from the same 
distribution. The torus region photons are thus significantly harder than
the point source (despite the 30\% contamination by the PSF wings); this
further supports the independant nature of this component.
We assume a PL model for this torus emission. Fixing $N_H$
as above and assuming a typical PWN $\Gamma = 1.5$, we find an unabsorbed flux 
$f_X({\rm 0.5-5~keV}) \approx 1.6 \pm 0.4 \times 10^{-14}{\rm erg/cm^2/s}$
for the full torus.  This spindown efficiency 
$\eta \approx 6 \times 10^{-5}d_{1.2}^2$ is about $10\times$ lower than 
that of most $\tau \sim 10^4$y pulsars.

\section{The SNR Association \& Kick-Spin Correlation}

	S147 is a late blast wave phase SNR, quite 
circular in both radio \citep{fr86} and H$\alpha$ \citep{vdb78} images, with 
a $\sim 166^\prime$ diameter. The estimated 0.8-1.6~kpc distance
agrees very well with the pulsar's 1.2~kpc DM-derived distance.
\citet{sfh80} find the geometrical center at 05:40, +27:44 
$(\pm 2.5^\prime )$ (J2000). At the nominal distance the $\gamma=5/3$ Sedov
solution gives a blast wave age of $t = 1.4 \times 10^5 d_{1.2}^{5/2}
(n/E_{51})^{1/2}$\,y and an expansion velocity $v_{\rm exp} = 82  
d_{1.2}^{-3/2} (E_{51}/n)^{1/2}$\,km/s for a local ISM density $n$ cm$^{-3}$
and an explosion energy $10^{51}E_{51}$erg. The optical
expansion velocity $\approx 80$\,km/s \citep{ka79} agrees well,
supporting the distance estimate. 
\begin{inlinefigure}
\scalebox{0.70}{\rotatebox{0}{\plotone{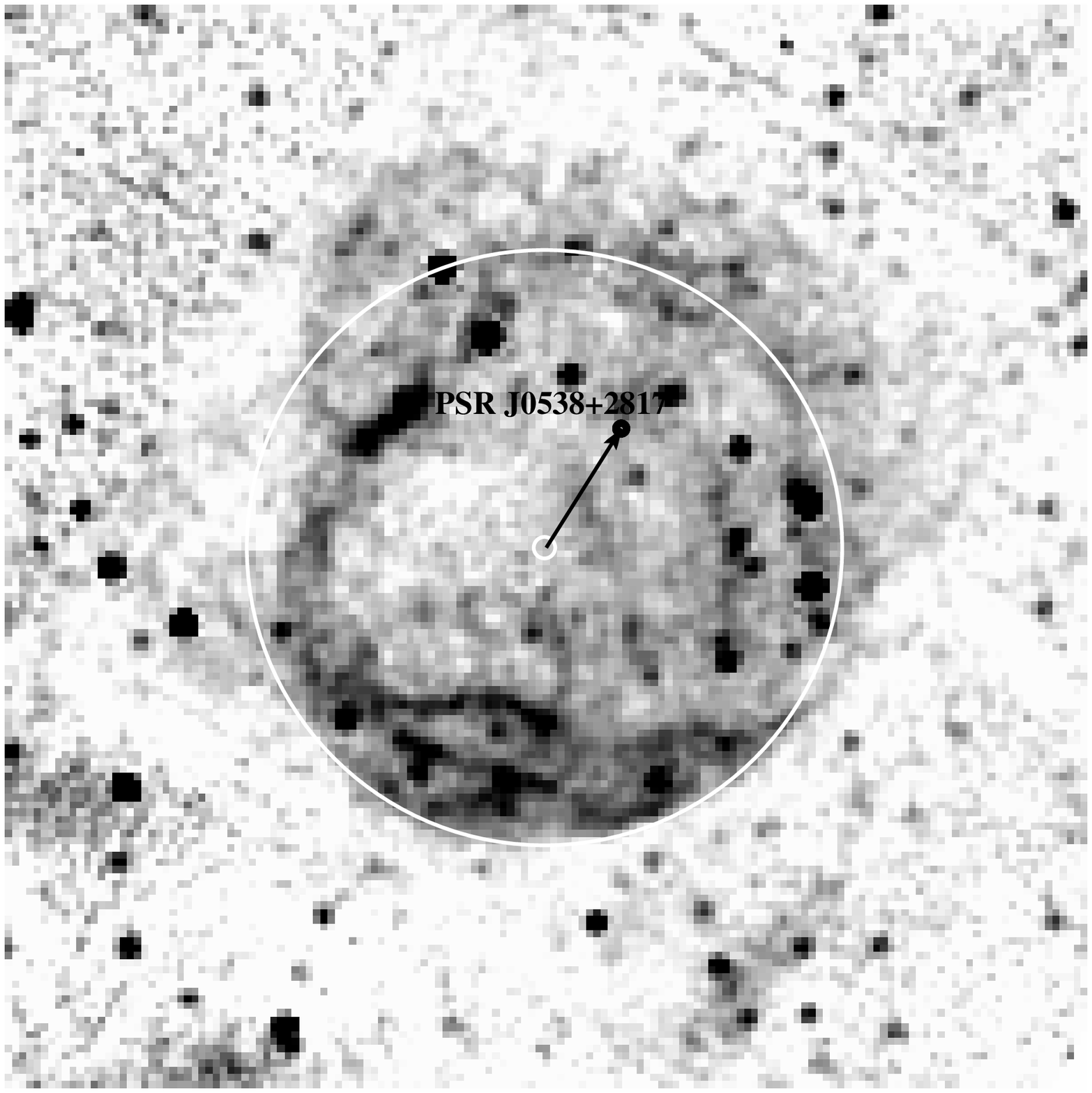}}}
\figcaption{
11cm image of S147 (after F\"urst \& Reich 1986), with \psr's proper motion 
vector (3$^\circ$ field, N up/E left).
\label{s147im}}
\end{inlinefigure}

If we assume magnetic dipole spindown ($n=3$) with
a constant field $10^{12}B_{12}$G, the spindown luminosity
at characteristic age $10^5\tau_5$y is 
$$
{\dot E} = 1.0 \times 10^{36} (B_{12} \tau_5)^{-2} {\rm erg/s},
$$
leading to a scale for the wind termination
$$
r_{T} \approx ({\dot E}/4\pi c P_{ext})^{1/2}.
$$
This termination will be azimuthally symmetric (e.g. toroidal) whenever
the external static pressure
$P_{ext} \ge P_{ram} = 6 \times 10^{-10}n v_7^2 {\rm g/cm/s^2}$, i.e.
subsonic pulsar motion at a speed $100v_7$km/s.  In the general ISM, $P_{ext}$
is typically a few$\times 10^{-13} {\rm g/cm/s^2}$; all but the
slowest young pulsars should display non-toroidal bow shocks.
However, in the interior of a Sedov-phase SNR we have 
$P_{ext} \approx 6 \times 10^{-11} (n_0^3E_{51}^2/t_5^6)^{1/5}{\rm g/cm/s^2}$
where the ambient ISM density is $n_0$ and we have distinguished between
the true (=SNR) pulsar age $t$ and the characteristic age $\tau_c$. Since the 
density in the hot
interior is quite low, eg. ${\rm Log}(n_i ) = {\rm Log}[n_0(2r/r_{\rm Sed}
-2)]$ \citep{ms74}, we expect pulsars in the interior of SNR to 
be in the  toroidal shock phase for $t_5 \la 3$. The difficulty is
that at typical speeds of $v_7 \approx 5$, a pulsar outruns the blast wave
by $t_5 \approx 0.3 (E_{51}/n_0)^{1/3}(v_7/5)^{-5/3}$. In fact, as soon as
it passes the contact discontinuity, the large density and ram pressure jump
ensure a bow shock morphology.  PSR 1951+32 in CTB 80
is likely in this phase (Shull \et\, 1988).

	According to these estimates, a torus for
\psr is not unexpected. At $t_5 \ga 1$ we would normally 
find it in the ambient ISM, but its low $v_7 \approx 1.5/t_5$ keeps it in its
host SNR. With a low $B_{12} = 0.74$, the pulsar husbands its
spin-down luminosity giving a torus of angular size
$$
\theta_{tor} = r_{T}/d \approx 14^{\prime\prime}(B_{12}d_{\rm kpc} \tau_5 
n_0^{0.3}E_{51}^{0.2}t_5^{-0.6})^{-1}
$$
or about $3^{\prime\prime}t_5^{0.6}n_0^{-0.3}d_{1.2}^{-1}$ for \psr,
among the largest angular sizes expected for known pulsars (after Crab
and Vela).
Since the observed inner edge of the torus is at $\sim 5^{\prime\prime}$,
these estimates agree if $t_5\sim 2$ or $n_0 \sim 0.2$.

	Given the large offset of \psr from the remnant center, a good
test of the ${\vec \Omega}$-${\vec v}$ correlation can be made. The offset vector
implies a proper motion $(23.6\pm 1.5)/t_5$\, mas/y at position angle
$-32.3\pm 3.6^\circ$. The pulsar transverse velocity is 134$d_{1.2}/t_5~$km/s.
Together with the torus geometry estimates of \S2.2, this offset
gives a kick-spin position angle difference of 
$\theta_{\Omega-v} = 6.3 \pm 6.6^\circ$.  If ${\vec v}$ is along the torus 
axis, the 3D velocity is $137 d_{1.2}/t_5$km/s, rather low for an isolated pulsar.

	The Crab pulsar optical proper motion \citep{cm99} 
and inner nebula axis vectors give $\theta_{\Omega - v} = 10 \pm 10^\circ$.
For Vela, the symmetry axis of the X-ray torus in \citet{pksg01} 
is aligned within $2\pm 3^\circ$ \citep{bet89} or $10\pm 3^\circ$ \citep{dmc00}
using the radio or optical ${\vec v}$, respectively. The joint 
probability of obtaining these (2-D) projections by chance is $\sim 0.3-1$\%, 
making the association intriguing, but not definitive.
If one adopt \psr as a third alignment, the chance probability 
drops to $\sim 2 \times 10^{-4}$. 

\section{Discussion: Age and Kick Constraints}

	One of the key discriminants of the origin of this correlation is
the initial pulsar spin period $P_0$ \citep{lcc01}. For
a constant braking index $n$ one finds
$$
t  = 2\tau_c  [ 1 - (P_0/P)^{(n-1)} ]/(n-1).
$$
Since our PWN position angle supports the connection with S147, we must
reconcile the $\tau_c = 10^{5.8}$y ($n=3$) characteristic age with the maximum
plausible age $\sim  10^5$y of the SNR. One possibility is very rapid
$B$ decay, giving an effective $n\ga10$. More likely, as for
several other PSR-SNR associations there is a large initial spin period.
The observed thermal luminosity supports the young age and large $P_0$.
This $L \approx 2 \times 10^{33}d_{1.2}^2 {\rm erg/s}$, interpreted
as full surface cooling agrees well with the flux in standard cooling
scenarios near $10^5$y.  
At $P_0 \approx 130$ms ($t_5 \approx 1$) \psr has one of the slowest 
initial spins known.

	Physical kick-spin connections generally rely on rotational
averaging \citep{sp98}, so this large $P_0$ is problematic for many proposed 
mechanisms. If the kicks are isotropically distributed, the net moment at
$\theta \approx 60^\circ$ must be averaged for $\tau_{\rm kick} \gg
0.38 P_0/{\rm tan}\theta_{\Omega-v}$ to achieve a final spin-kick angle 
$\theta_{\Omega-v}$.
Indeed, since `Natal' kicks may be applied before the neutron
star has reached its final $\sim 10^6$cm radius, conservation of
angular momentum for kicks at $r_6 > 1$ increases the effective initial 
period by $r_6^2$. Thus for \psr, we require
$\tau_{\rm kick} \gg 0.5\;{\rm s} (1-t_5/6)^{1/2}r_6^2$
for the observed $\theta_{\Omega-v}$.

Certainly the E-M rocket (Harrison-Tademaru) effect is hopeless,
as it requires $P_0 \la 3$ms for 100km/s velocities. Also hydromagnetic
models imparting momentum over $\sim 0.1$s are inadequate, especially as
the momentum deposition is at the $r_6 \ga 10$ shock break-out.
Even mechanisms invoking asymmetric $\nu$ emission are severely constrained,
but appear just feasible considering the $\sim 3$s $\nu$ diffusion timescale at 
$r_6 \sim 1-2$. However
inducing the large required $\nu$ asymmetry is challenging. For 
magnetic-field induced asymmetry, even \psr's modest velocity requires
an ordered neutrinosphere field of $\sim 10^{15.5}$G \citep{al99}, otherwise
unmotivated for this low $B$ pulsar.

	As an alternative one may imagine schemes where the spin is driven
to the kick axis. One possibility is post-collapse accretion of a remnant
disk (Blandford, private comm.). Imagine a disk of $\sim \Delta M$ accreting 
at a super-Eddington rate onto a $1.5M_{1.5}M_\odot$ neutron star with 
moment of inertia $10^{45} I_{45}{\rm g cm^2}$. The disk is approximately 
Keplerian down to some inner radius $r_{in}$, at which point we assume
roughly half of the mass accretes while the remainder is ejected
in polar jets with velocity $v_{esc} \approx 2^{1/2} v_{\rm Kep}(r_{in})$ and
an asymmetry $\chi$. The large $P_0$ inferred for \psr is a 
serious constraint on this scenario. If the disk extends to the star, we 
might expect jet velocities of $\sim c/4$ as in SS443, but only 
$\Delta M \approx 2\times 10^{-3} I_{45}(100{\rm ms}/P_0)\; M_\odot$ 
may accrete before one spins up to $P_0< 0.1$s.
With this mass, jet recoil gives a kick
$v_K \approx 75 \chi/M_{1.5}$km/s for \psr's parameters, so that even a large
jet asymmetry does not produce the desired momentum.

	Perhaps more realistically, standard equilibrium spin-up arguments set 
$r_{in} = (B_\ast^2 R_\ast^6/{\dot M} \sqrt{2\,G\,M_\ast})^{2/7} 
\approx 1.8 \times 10^8 B_{12}^{4/7}{\dot M}_{-8}^{-2/7}$cm. The Keplerian
period
at this Alfven radius is thus comparable to $P_0$ for ${\dot M}_{-8} \approx
130$, i.e. $\sim 100\times$ the Eddington accretion rate. 
If as expected the star is born near a $\sim 1\,P_i$ms break up period,
the minimum mass needed to spin down the star to $P_0$ is
$\Delta M \approx 0.04 M_\odot I_{45}P_i^{-1}M_{1.5}^{-2/3}
(P_0/100{\rm ms})^{-1/3}$,
which gives a kick velocity
$v_k \approx 800 \chi I_{45}M_{1.5}^{-1/3}(P_0/100{\rm ms})^{-2/3}/P_i$~km/s.
\psr is relatively slow with $v \approx 140$~km/s and $P_0 \approx 130$ms, so
we require an jet asymmetry $\chi \approx 0.2 P_i M_{1.5}^{1/3}/I_{45}$.
Since the velocities are set at $r_{in}$ we see that slow spin goes
with slow kick in this model. Substantial jet asymmetries are
required, but remain to be explained. Of course, additional mass accreted 
while at equilibrium reduces the required $\chi$.

	With its large $P_0$ and small $\theta_{\Omega-v}$, \psr provides 
some of the strongest constraints on spin-kick physics. These important physics
conclusions are presently based on rather limited data; clearly
better images confirming the torus and refining
$\Psi$ are needed. Also, a radio interferometric proper 
motion can improve the velocity vector and get an independent kinematic measure
of $t_5$.  Such data should provide a new window on neutron star
core collapse and its immediate aftermath.

\acknowledgments

We thank the referee, D. Sanwal for a careful reading and useful recommendations.
This work was supported by CXO grant G02-3085X.


\begin{deluxetable}{cccccc}
\tabletypesize{\normalsize}
\tablecaption{Point Source Spectral Fits\label{tbl-pf}}
\tablehead{
Model & kT(keV) & $N_H$($10^{21}{\rm cm^{-2}}$) & 
$f_x$($10^{-12}{\rm erg/cm^2/s}$) & $R_{eff}$(km) & $\chi^2$/DOF\\
}
\startdata
BB & 0.159$\pm 0.0017$ & 3.1$\pm0.2$ & 3.3 & 2.6 & 1.03 \\
magH  & 0.054$\pm 0.0007$ & 4.7$\pm0.2$ & 13.7 & 13 & 0.94 \\
\enddata
\end{deluxetable}

\end{document}